# Initiation and blocking of the action potential in the axon in weak ultrasonic or microwave fields

M. N. Shneider[1],[*] and M. Pekker[2]

[1] Department of Mechanical and Aerospace Engineering, Princeton University, Princeton, NJ 08544, USA

[2] Drexel Plasma Institute, Drexel University, 200 Federal Street, Camden, NJ 08103, USA

**Abstract**
In this paper, we analyze the effect of the redistribution of the transmembrane ion channels in the axon caused by longitudinal acoustic vibrations of the membrane. These oscillations can be excited by an external source of ultrasound and weak microwave radiation interacting with the charges sitting on the surface of the lipid membrane. It is shown, using the Hodgkin-Huxley model of the axon, that the density redistribution of transmembrane sodium channels may reduce the threshold of the action potential, up to its spontaneous initiation. At the significant redistribution of sodium channels in membrane, the rarefaction zones of the transmembrane channels density are formed, blocking the propagation of the action potential. Blocking the action potential propagation along the axon is shown to cause anesthesia in the example case of a squid axon. Various approaches to experimental observation of the effects considered in this paper are discussed.

1. Introduction

In the classical works of Hodgkin and Huxley [1,2,3], they have shown that the transmission of the nerve signal is associated with the action potential. In the process of the action potential along the axon membrane, there is a local change in the charge of the membrane surfaces and their vicinities (internal and external) to the opposite, which is then recovered during the so-called refractory period. The resting potential (~ (-70) mV) between the outer and inner surfaces of the membrane is supported by the work of the ion (sodium, potassium and calcium) channels (pumps). The transmembrane protein ion channels (Fig.1) are not fixed in any specific locations of the membrane and may move along its surface fast enough (the corresponding coefficients of lateral diffusion lie within $D_L = 10^{-13} - 10^{-14}$ m²/s [4, 5, 6]). It was shown in the paper [7], on the basis of the analysis of the equations of Hodgkin and Huxley [3], that the local increase in the density of the protein transmembrane sodium ion channels leads to a decrease in the excitation threshold of the action potential. The redistribution of ion channels leads to the appearance of the regions with high density of channels and the areas with ion channels depletion. These areas (as demonstrated in this paper) can block the propagation of action potentials along the axon.

The transfer of protein transmembrane ion channels along the membrane may be caused by various factors. In [8], the mechanism, associated with the acoustic pressure on the ion channels of the longitudinal ultrasonic (running or standing) wave excited in the membrane of the axon, was considered. A possible source of the forced oscillations of the membrane is the interaction of external microwave electromagnetic field with the charged surfaces of the axon's membrane, rather, the initial segment of the axon (the area between the neuron hillock and the first section of an axon covered with the myelin layer (Fig. 2)).

---

[*] m.n.shneider@gmail.com

Note that in [7], the mechanisms that could lead to the redistribution of ion channels were not considered. In [8], we considered how ion channels could be redistributed along the axon as a result of the acoustic pressure on the ion channels resulting from the longitudinal ultrasonic (running or standing) wave excited in the membrane by microwave fields. It has been shown that the longitudinal projection of the microwave electric field acting on the charges attached to the membrane leads to longitudinal deformation of the membrane which, in turn, results in the interaction with the transmembrane channels. This leads to a significant redistribution of the ion channels in the initial segment of the axon (the area between the neuron hillock and the first section of an axon covered with the myelin layer (Fig. 2)), without a noticeable increase in the local temperature near the neural fibers.

The role of anesthetics is to change the conductivity of the transmembrane of ion channels (see, for example, the review [9]). It was shown [10, 11, 12], that the action of anesthetics is linearly related to their solubility in membranes. It is well known that the anesthesia effect can be achieved without anesthetics using ultrasound. This method of anesthesia by ultrasound has long been known (see for example [13-14]). The traditional ultrasonic effect of anesthesia is associated with the irreversible impairment of the membrane integrity of nerve fibers, due to the development of cavitation in the regions irradiated by ultrasound [15].

A natural question arises: is it possible to reverse the blocking of the action potential (anesthesia) through the redistribution of transmembrane channels? That is, the reversible effect of the anesthesia is possible when, after switching off an ultrasound source, the equilibrium density of the protein ion channels per unit area is restored by the lateral diffusion.

In this paper, the influence of traveling or standing ultrasound waves on the protein channels redistribution is considered. It was shown that the thermal effects on the thresholds of the action potential become dominant when increasing the ultrasound intensity. On the other hand, as shown in [8], a significant redistribution of the ion channels' density (without noticeable change in membrane temperature) may be caused by the electro-acoustical vibrations arising in the membrane under the action of the weak microwave fields at specific resonant frequencies. This paper shows that this effect can lead to the reversible action potential blocking (reversible anesthesia).

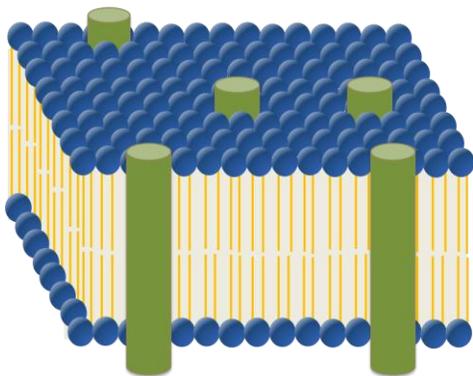

**Fig. 1**. An element of the lipid bilayer membrane of the axon with protein transmembrane channels

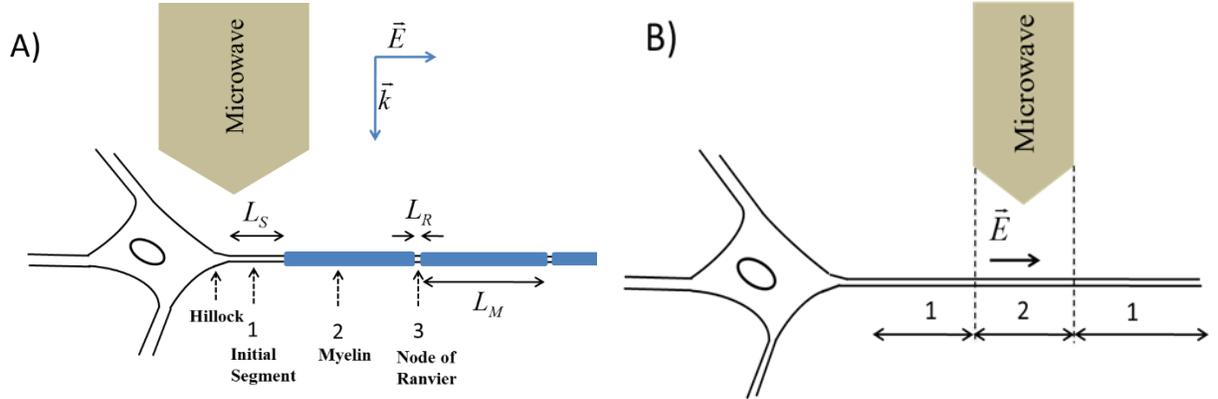

**Fig. 2.** A) Scheme of a neuron with the characteristic elements of the myelined axon. $L_s \approx 10-50\,\mu m$, $L_R \approx 1-2.5\,\mu m$, $L_M \approx 100-1000\,\mu m$. B) Scheme of a squid neuron. The length of axon 0.5- 5 m. 1 – the areas of axon, which are not affected by the electric field of the incident microwave field, the membrane is fixed. 2 – the axon area ($l \sim 5 - 10$ cm) irradiated by the microwave; the membrane is displaced by the force $\mathbf{F}=\sigma\mathbf{E}$.

In part II of this paper, the redistribution of the transmembrane protein ion channels as a result of the acoustic pressure is described. Both cases of running and standing waves are considered.

In the part III of the article, the amplitude of the ion channel redistribution at which the self-excited action potential in the axon starts, is found. The conditions for the blocking of the action potential along the axon are shown as well. All calculations are made on the basis of the standard model of Hodgkin and Huxley for the squid axon [3]. This model was chosen for the following reasons: it is well studied; the experiments with action potential self-excitation or blocking in a squid axon can be carried out relatively easily; and, most importantly, the squid axon is non-myelinated. That is, the numerical and experimental results will be valid for the initial segments in neurons, where the action potential is initiated.

In the part IV, the possibility for experimental observation of self-excitation of the axon and of action potential blocking (effect of anesthesia) in the squid axon is considered.

In the part V, the obtained results are discussed, and conclusions are presented.

## II. The mechanism of redistribution of transmembrane protein ion channels associated with the acoustic pressure on the ion channels

The transfer of sodium ions in the nerve fibers is carried out through transmembrane protein channels permeating the bilayer membrane (Fig. 1). These channels are located in the axon's initial segment and the nodes of Ranvier (Fig. 2). Under the normal conditions, the density of channels per the membrane unit area, despite their mobility, is practically constant and does not depend on the action potential excited in the axon or not. It is shown in [8] that the transmembrane channels redistribution in the axon can occurs as a result of interaction with a standing or traveling ultrasonic wave. It is also shown in [8] that the source of the acoustic standing wave can be microwave electric field wave in GHz's frequency range. However, a narrowly focused ultrasound antenna can serve as a source of excitation of acoustic waves in the membrane as well.

Not considering the source of excitation of longitudinal sound waves in the membrane, we estimate the force acting on the sodium channel in the cases of standing or traveling wave.

For simplicity, we assume that the sodium channel is not a cylinder, but a sphere of radius $a = h = d/2$, where $d$ is the thickness of the membrane and $a < \lambda$ ($\lambda$ is the ultrasound wave length). For numerical estimates we will apply a simple analytical formulas derived for spherical particles in an ultrasonic field. We hope that for real channels (of cylindrical shape) results will not be much different.

In the case of a traveling ultrasound waves, ultrasound waves are scattering by macro particles and transfer momentum to them, therefore force, acting on spherical macro particle radius $a$, is [16]:

$$F_t = 4\pi a^6 k^4 \frac{1 + \frac{2}{9}(1-\varsigma)^2}{(2+\varsigma)^2} \overline{E} . \tag{1}$$

In the case of a plain standing acoustic wave, the amplitude of the waves increases with increasing distance from the node. Thus the rms pressure acting on the macroparticle from the site of the node is lower than from the site of the maximum amplitude, and the particle is shifted toward the node. Accordingly, the force acting on the sphere of the radius $a$ is [16]:

$$F_s = 2\pi a^3 k_n \overline{E} \sin(2k_n x)[1 + \tfrac{2}{3}(1-\varsigma)]/(2+\varsigma) . \tag{2}$$

Here, $\overline{E} = \frac{1}{2}\rho_0 \omega^2 A$ is the energy density in the wave; $\omega = 2\pi f$ is the angular frequency of the ultrasound wave; $A$ is the amplitude of the oscillatory displacements in the wave; $k = 2\pi/\lambda = c_s/\omega$ is the wave number; $\varsigma = \rho/\rho_a$, $\rho, \rho_a$ are the densities of the medium and the particles; $x$ is the position of a particle in the standing wave, $k_n = \frac{n\pi}{l}$, $n = 1,2,3...$, $l$ is the size of the area where a standing wave exists.

For the case $\varsigma = \rho/\rho_a \approx 1$, (1) and (2) are reduced, correspondingly:

$$F_t = \frac{4\pi}{9} a^6 k^4 \overline{E} \tag{3}$$

$$F_s = \frac{2\pi}{3} \cdot a^3 k_n \overline{E} \sin(2k_n x) \tag{4}$$

Under the action of acoustic radiation (i.e. pressure force (3) or (4)), the transmembrane proteins channels are displaced along the $x$-axis. After the termination of these forces, the equilibrium distribution of ion channels is restored due to lateral diffusion.

Here in after, we follow [8]. The continuity equation for the transmembrane protein channels is as follows:

$$\frac{\partial n_{ch}}{\partial t} + \nabla \cdot \mathbf{\Gamma} = 0; \quad \mathbf{\Gamma} = \mu_L \mathbf{F} n_{ch} - D_L \nabla n_{ch} \tag{5}$$

Here, $n_{ch}$ is the density of the protein channels per unit membrane surface; $D_L$ is the lateral diffusion coefficient; $\mu_L$ is the lateral mobility of transmembrane protein channels, which is associated with the lateral diffusion and the local temperature by the Einstein relation:

$$\mu_L = D_L / k_B T .\tag{6}$$

We assume that the external force **F** associated with the pressure of a plane ultrasonic wave on the transmembrane channels is directed only along the axis of the axon. Since the membrane thickness is much smaller than the radius of the axon, the membrane can be considered as a flat plate, unbounded in a direction perpendicular to the wave propagation. In this case, equation (5) can be rewritten as

$$\frac{\partial n_{ch}}{\partial t} = D_L \frac{\partial}{\partial x}\left(\frac{\partial n_{ch}}{\partial x} - \frac{F}{k_B T} n_{ch}\right).\tag{7}$$

When reaching the equilibrium, the drift flux of protein transmembrane channels is compensated by the lateral diffusion, therefore:

$$\frac{\partial n_{ch}}{\partial x} - n_{ch}\frac{F}{k_B T} = 0 .\tag{8}$$

Thus, the steady-state distribution of the protein channels density satisfies the Boltzmann distribution:

$$n_{ch}(x) = n_0 \exp(-U(x)/k_B T)\tag{9}$$

with the potential

$$U(x) = -\int_0^x F(x)d\,x .\tag{10}$$

The factor $n_0$ can be easily found from the conservation condition of the total number of ion channels along the length of the membrane.

$$n_{ch}(x) = \frac{n_{ch,0}}{\frac{1}{l}\int_0^l e^{-U(x)/k_B T} dx} e^{-U(x)/k_B T} ,\tag{11}$$

here $n_{ch,0}$ is the undisturbed density of ion channels per unit membrane surface. Consider the influence of ultrasonic waves on the initial segment (Fig. 2). The dimensions of the initial segment $L_s$ vary between 10-50 μm, which is much smaller than the wavelength of ultrasonic vibrations. Since the initial segment is bordered on one side with the myelin covered area, and, on the other, with hillock, the longitudinal displacement of the ends of the initial segment membrane is suppressed. Thus, only standing waves with wave numbers $k_n = \pi n / L_s$, $n = 1,2,3\ldots$ can exist in the membrane of the initial segment (Fig. 3).

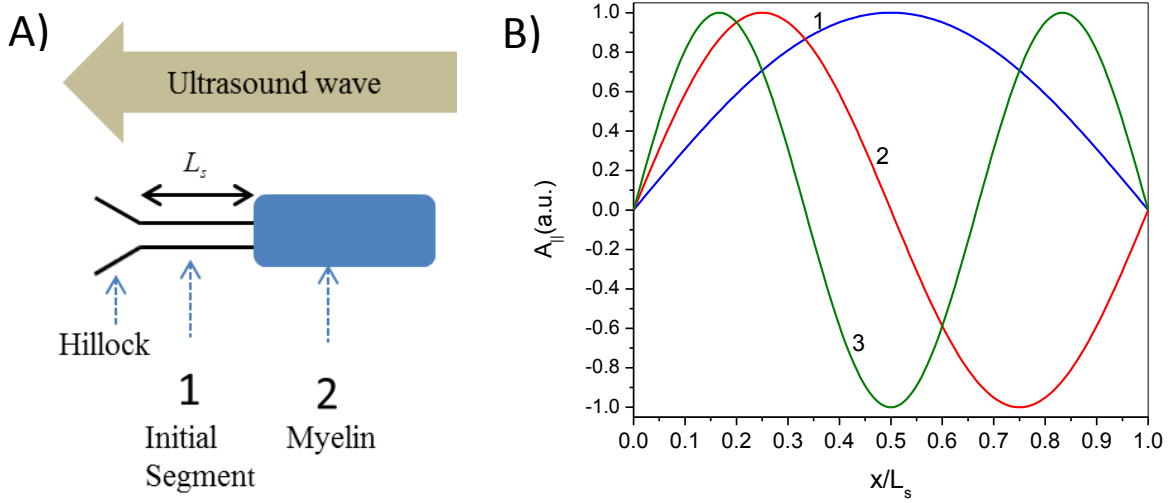

**Fig. 3.** Schematic view of ultrasonic waves acting on the initial segment. Fig. A) A general view, Fig. B) The longitudinal displacement amplitude in the membrane; line 1 corresponds to $k_1 = \pi/L_s$, 2 - $k_2 = 2\pi/L_s$, line 3 - $k_3 = 3\pi/L_s$, Fig. B) correspondc to the initial segment.

For the case of a standing wave from the formula (11) follows:

$$\frac{n_{s,ch}}{n_{ch,0}} = \frac{e^{-\frac{\pi}{3}a^3 \overline{E}_s (1-\cos 2k_n x)/k_B T}}{\frac{1}{L_s}\int_0^{L_s} e^{-\frac{\pi}{3}\cdot a^3 \overline{E}_s (1-\cos 2k_n x)/k_B T} dx} = \frac{e^{\frac{\pi}{3}\frac{a^3 \overline{E}_s}{k_B T}\cos(2k_n x)}}{I_0\left(\frac{\pi}{3}\cdot\frac{a^3 \overline{E}_s}{k_B T}\right)}, \tag{12}$$

Where $n_{s,ch}$ is the density of the ion channels in the case of the standing ultrasound wave, $\overline{E}_s = \frac{1}{2}\rho_0 \omega^2 A_s^2$ is the energy density of the standing ultrasonic wave, $I_o$ is the modified Bessel function [17]. Figure 4shows the dependence of the equilibrium density of sodium channels along the initial segment. The difference between the maximum and the minimum channel density is:

$$\frac{\Delta n_{s,ch}}{n_{ch,0}} = \frac{e^{-\frac{\pi}{3}a^3 \overline{E}_s (1-\cos 2k_n x)/k_B T}}{\frac{1}{L_s}\int_0^{L_s} e^{-\frac{\pi}{3}\cdot a^3 \overline{E}_s (1-\cos 2k_n x)/k_B T} dx} = \frac{2\sinh\left(\frac{\pi}{3}\cdot\frac{a^3 \overline{E}_s}{k_B T}\right)}{I_0\left(\frac{\pi}{3}\cdot\frac{a^3 \overline{E}_s}{k_B T}\right)} = \frac{2\sinh(\xi_s)}{I_0(\xi_s)}, \qquad \xi_s = \frac{\pi a^3 \overline{E}_s}{3k_B T} \tag{13}$$

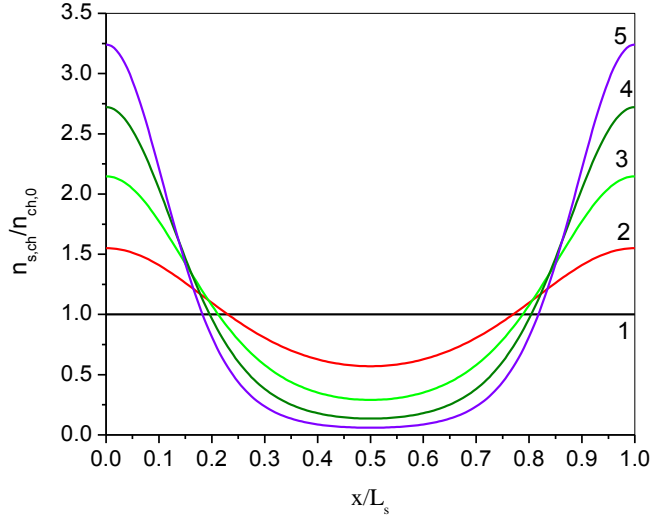

**Fig. 4.** The relative densities of sodium channels along the initial segment interacting with the main mode of an ultrasound standing wave ($n=1$) for different values of the parameter $\xi_s$. Line 1 corresponds to $\xi_s = 0$, 2 - $\xi_s = 0.5$, 3 - $\xi_s = 1$, 4 - $\xi_s = 1.5$, 5 - $\xi_s = 2$.

Note that, at $\xi_s > 1$, the sodium channels are almost completely shifted to the edges of the initial segment. On the one hand, this may lead to the blockage of the action potential. On the other hand, this may reduce the excitation threshold of the action potential, as there are areas in which the channel density is much higher than at the unperturbed equilibrium. Examples of forced redistribution of transmembrane channels which lead to the blockage of the action potential and to self-excitation are considered in the following sections.

Figure 5 shows the density of sodium channels at different moments in time obtained through numerical solution of equation (7). It can be seen that the channels redistribution happens quite quickly. At $\xi_s = 2$, it takes approximately 5% of the characteristic time for the lateral diffusion, $t_0 = L_s^2 / D_L$.

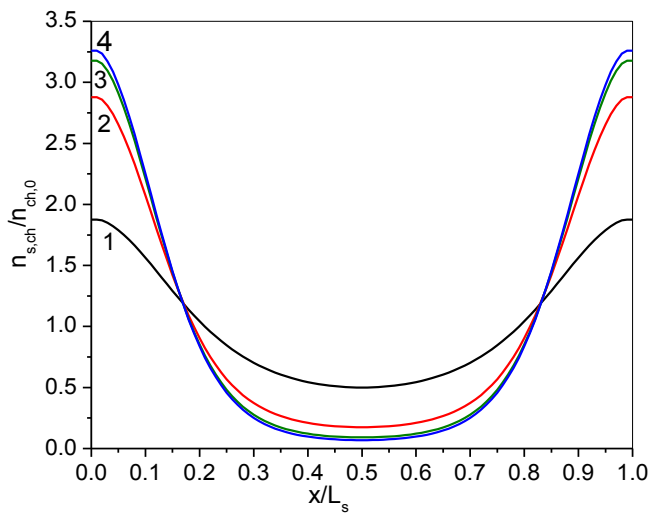

**Fig. 5.** The relative densities of sodium channels along the initial segment interacting with the main mode of ultrasound standing wave ($n=1$) for parameter $\xi_s = \pi a^3 \overline{E}_s / 3k_B T = 2$ at different

moments in time. 1 corresponds to $t/t_0 = 0.01$, ($t_0 = L_s^2/D_L$), $2 - t/t_0 = 0.03$, $3 - t/t_0 = 0.05$, $4 - t/t_0 = 0.07$.

It follows from Fig. 5 that for $\xi_s = 2$, the formation of a new equilibrium distribution of ion channels happens very quickly in the time interval of order $0.03 t_0$. In other words, the initial formation stage of the regions with reduced or increased density of ion channels is much shorter than the channel density restoration time after turning off the ultrasound.

Figure 6 shows the attenuation length dependence on the sound frequency. The dimensions of the initial segment vary between 10-50 µm. Since the decay length must be greater than the segment's length, the respectively allowed frequencies of the ultrasonic waves are in the range 1-20 GHz.

For the case of a traveling wave, we must take into account that the density of energy of the acoustic waves are decaying due to the viscosity as:

$$\overline{E}_t = \overline{E}_{t,0} \exp(-2x/a_d), \tag{14}$$

where $a_d = c_s^3/(\omega^2 \nu)$ is the attenuation length of sound waves in water, $\nu \approx 10^{-6}/(1+\omega^2 \tau_\nu)$ m²/s is the kinematic viscosity, $\tau_\nu \approx 8 \cdot 10^{12}$ s [18], $c_s = 1450$ m/s is speed of sound in the membrane, which we assumed the same as for water. Figure 6 shows a sound attenuation length dependency on the frequency in water.

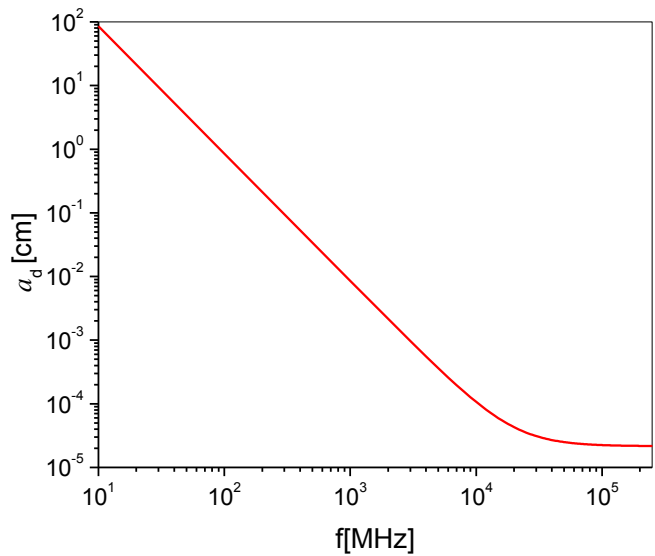

**Fig.6.** Sound attenuation length versus frequency in water

The solution of equation (7) for a traveling wave ($l \gg a_d$) is shown in Figure 7. In the calculations, we have selected the function $\dfrac{F}{k_B T}$, which is included in equation (7), as follows:

$$\frac{F}{k_B T} = \frac{\xi_t}{a_d}\exp(-2x/a_d) \text{ at } x \geq 0 \text{ and } \frac{F}{k_B T} = \frac{\xi_t}{a_d}\exp(6x/a_d) \text{ at } x < 0.$$

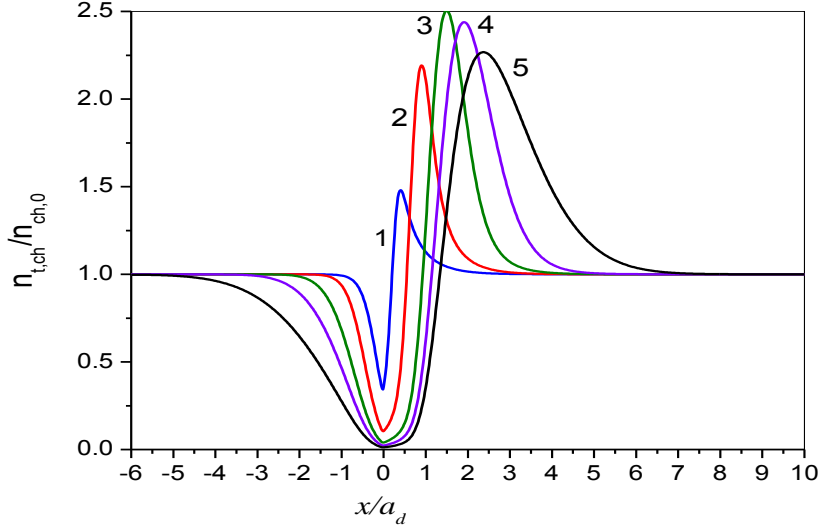

**Fig.7.** Dependence of surface density of sodium channels along the length of axons in the case of an ultrasonic traveling wave. $t_0 = a_d^2/D_L$, $\xi_t = 40$. Curve 1 corresponds to the $t/t_0 = 0.01$, ($t_0 = a_d^2/D_L$), $2 - t/t_0 = 0.05$, $3 - t/t_0 = 0.2$, $4 - t/t_0 = 1$, $5 - t/t_0 = 1.5$.

It follows from Fig. 7 that at $Fa_d/k_B T = 40$ the cannels' surface density is decreasing rapidly, for the time of the order $0.01 t_0$, but the formation of the depleted with the ion channels region $l \approx a_d$ happens in the time period of the order $t \approx t_0$.

It is noteworthy that if the ultrasound is directed towards the hillok (Fig. 2A), it is possible to block the signal because the action potential cannot jump over an area with the essentially reduced density of ion channels. However, if the ultrasound is directed from the hillock, then a high density of sodium channels appears in front of the first myelin covered area, which reduces the excitation threshold and may even lead to the self-excitation of the action potential.

### III. The initiation and blocking of the action potential in an axon.

As shown in the preceding paragraph, the ultrasound can lead to a redistribution of sodium channels. As a result, regions of high and low transmembrane sodium channels density may be formed (Fig. 4, 5, 7). Since the generation of an action potential in a neuron happens in the initial segment (i.e. the non-myelinated part of the axon between the axon hillock and the first myelinated section) all the calculations below were done on the basis of the standard model of Hodgkin and Huxley for the squid axon [3]. This model was chosen for the following reasons: it is well studied; the experiments on self-excitation or blocking of the action potential propagation in the axon of the squid can be carried out with relatively ease; and the squid's axon is non-myelinated. That is, the numerical and experimental results will also be qualitatively valid for the initial segments in neurons. The system of equations of Hodgkin and Huxley has the form [3]:

$$\frac{C}{r_m c_m} \frac{\partial^2 V_m}{\partial x^2} - C \frac{\partial V_m}{\partial t} = G_K \cdot n^4 \cdot (V_m - V_K) + G_N \cdot m^3 h \cdot (V_m - V_{Na}) + G_L \cdot (V_m - V_L) \tag{17}$$

$$\frac{dn}{dt} = \alpha_n \cdot (1-n) - \beta_n n \tag{18}$$

$$\frac{dm}{dt} = \alpha_m \cdot (1-m) - \beta_m m \tag{19}$$

$$\frac{dh}{dt} = \alpha_h \cdot (1-h) - \beta_h h \tag{20}$$

$$\alpha_n = \frac{100 - 10u}{\exp(1 - 0.1u) - 1} \qquad \beta_n = 125 \cdot \exp\left(\frac{u}{80}\right) \tag{21}$$

$$\alpha_m = \frac{2500 - 100u}{\exp(2.5 - 0.1u) - 1} \qquad \beta_m = 4000 \exp\left(\frac{u}{18}\right) \tag{22}$$

$$\alpha_h = 70 \cdot \exp\left(\frac{u}{20}\right) \qquad \beta_h = \frac{1000}{\exp(3 - 0.1u) - 1} \tag{23}$$

Here, $V_m$ is the membrane potential in millivolts, $V_K = -77\,\mathrm{mV}$, $V_{Na} = 50\,\mathrm{mV}$, $V_L = -54.4\,\mathrm{mV}$ $u = V_m - V_R$, $V_R = -65\,\mathrm{mV}$ is the equilibrium potential difference (resting potential) on the membrane, $r_m = 2 \cdot 10^4\,\Omega/\mathrm{cm}$, $c_m = 1.5 \cdot 10^{-7}\,\mathrm{F/cm}$, $C = 10^{-6}\,\mathrm{F/cm^2}$, $G_K = 0.036\,\Omega^{-1}\mathrm{cm}^{-2}$, $G_{Na}^0 = 0.12\,\Omega^{-1}\mathrm{cm}^{-2}$, $G_L = 3 \cdot 10^{-4}\,\Omega^{-1}cm^{-2}$. $G_{Na}^0$ is the unperturbed conductivity for sodium ions per unit surface of the axon membrane. In general, the coefficients (21) - (23) depend on the temperature. Without loss of generality, all calculations are made for the temperature 6.3 C, as in [3].

The redistribution of sodium channels was specified in calculations by the model function (Fig. 8), so that the distribution of the sodium surface conductivity is $G_{Na} = G_{Na}^0 \cdot (1 + \xi_{Na}(x))$. The parameters $\xi_{Na}, \xi_0$ characterize the perturbation of the sodium channels density and the amplitude of this perturbation. The length of the model axon was equal $l_a = 40$ cm.

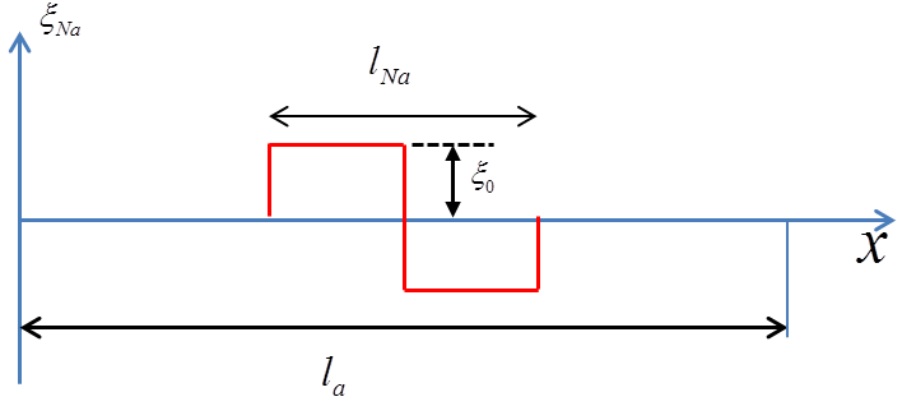

**Fig.8.** The model function $\xi_{Na}$ and the size of the redistribution channels, $l_{Na}$.

The excitation of the axon in the model was determined by a local increase in the potential in the point $x_0 = l_a/2$ for the time interval $\tau_0 = 1$ ms:

$$V_0(x_0) = V_R + \delta V \theta(t-\tau_0) = V_R + \alpha |V_R| \theta(t-\tau_0), \qquad (24)$$

where $\theta(t-\tau_0) = \begin{cases} 1, & t \leq \tau_0 \\ 0, & t > \tau_0 \end{cases}$ is the step function.

Here, the parameter $\alpha$ characterizes the amplitude of the action potential perturbation. When $\alpha = 0$, the perturbing potential $\delta V = 0$, and the potential at the point $x_0$ is equal to the unperturbed potential of the membrane. The dependence of the parameter $\alpha$ on $\xi_0$ ($l_{Na}=9$ cm, $x_0 = l_a/2 = 20$ cm) is shown in Fig.9.

It is seen that when increasing the parameter $\xi_0$, the threshold value for the action potential excitation decreases, and the self-excitation takes place in the axon at $\xi_0 = 0.331$.

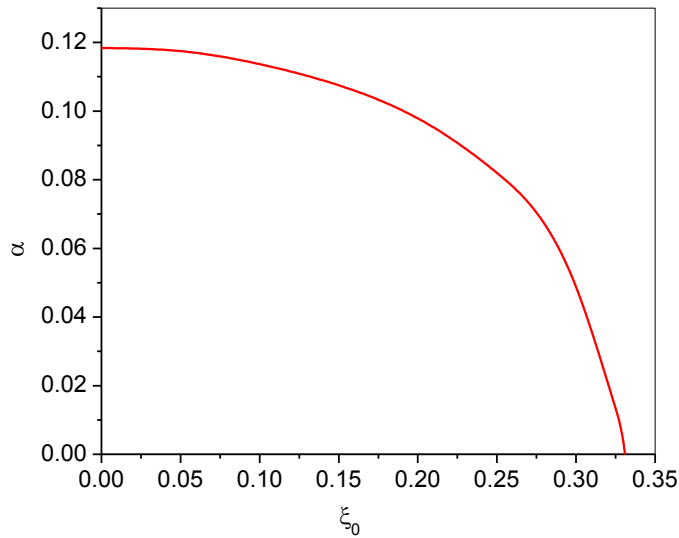

**Fig.9.** Parameter $\alpha$ on $\xi_0$ at $l_{Na} = 9$ cm, $x_0 = 20$ cm.

The time dependence of the action potential in the axon at different positions for the case $\xi_0 = 0.3$ [5] is shown in Fig. 10. The spontaneous excitation of the potential takes place at $x = 20$ cm, and spreads farther in both directions. It also follows from Fig. 10 that the velocity of the action potential is $v_f \approx 12.5$ m/sec, and the rise time is of the order of $\tau_p \approx 3$ ms.

Let's estimate a spatial scale of the action potential $l_x$. It follows from (17), that the diffusion coefficient of the action potential $D \approx 1/r_m c_m$. For the known velocity of the pulse front $v_f$, the estimate follows:

$$l_x \approx D/v_f = 1/c_m r_m v_f \approx 0.25 \text{ cm} \tag{25}$$

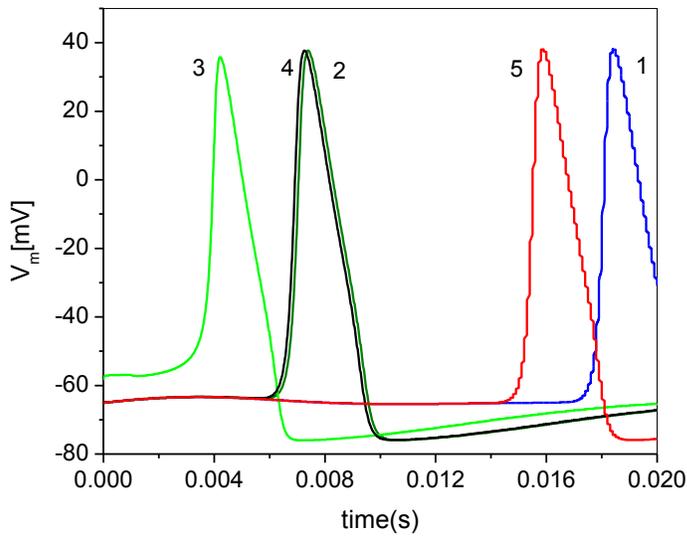

**Fig.10.** The time dependence of the action potential at various points in the axon. The excitation potential is 0, $\xi_0 = 0.3$ [5]. Line 1 corresponds to $x = 2$ cm, 2- 15.5 cm, 3– 20 cm, 4- 24.5 cm, 5– 35 cm.

By substituting the values $v_f$, $c_m$, $r_m$ in (25) we find $l_x \approx 0.25$ cm, that is in agreement with the numerical simulations shown in Fig.11.

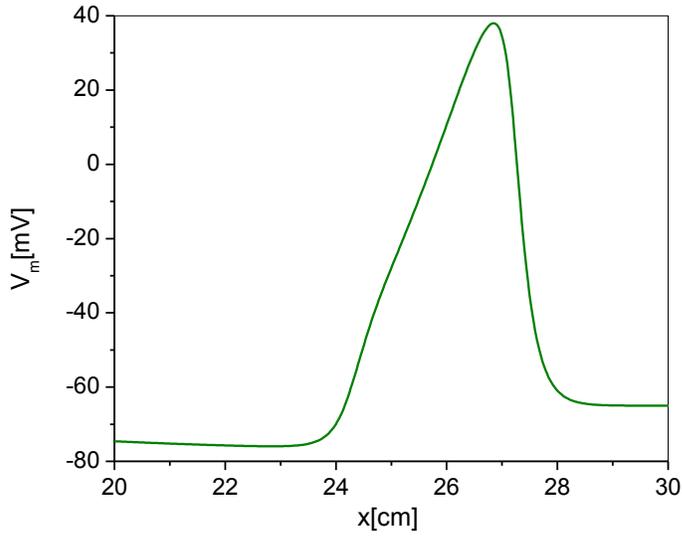

**Fig.11.** The instantaneous spatial structure of the action potential.

The reason why the density redistribution of transmembrane proteins leads to a lowering of the excitation threshold of the action potential is as follows: the change in the density of sodium transmembrane ion channels is equivalent to the appearance of additional local current density associated with the deviation of the surface density of sodium channels from the equilibrium distribution in region $l_{Na}$:

$$\delta j = (G_{Na} - G_{NA}^0) \cdot m_0^3 h_0 \cdot (V_R - V_{Na}) \quad (26)$$

Here, $m_0 = \alpha_m /(\alpha_m + \beta_m) = 0.053$ and $h_0 = \alpha_h /(\alpha_h + \beta_h) = 0.6$ are the equilibrium values, $m$ and $h$ are described by the equations (18) and (19), respectively. The estimate of the perturbation of membrane potential $\delta V_m$, which corresponds to the current density perturbation $\delta j$, follows from equation (17):

$$\frac{1}{V_m}\frac{\partial V_m}{\partial t} \approx \frac{1}{V_m}\frac{\delta V_m}{\tau_f} = \frac{1}{V_m}\frac{\delta j}{C} \quad (27)$$

When substituting the value $\delta j$, given by (26), into (27), and taking into account that the difference in the protein channels surface density in the interaction region is equal to $2\xi_0$ (Fig. 9), we obtain:

$$\frac{\delta V_m}{V_R} \sim \frac{2\tau_p}{CV_R}\xi_0 G_{NA}^0 m_0^3 h_0 (V_R - V_{Na}) \approx 0.06\xi_0 \quad (28)$$

By increasing the parameter $\xi_0$, the conditions of propagation to the right, i.e. in the region of low density of ion channels in the signal, deteriorates, and at $\xi_0 = 0.787$, the action potential does not extend to the right, i.e., in this direction, the blocking of the axon takes place. Figure 12 shows the dependence of the action potential on time at $\xi_0 = 0.8$ [1]. In this case, at $x = 20$cm the spontaneous excitation of the action potential propagating to the left occurs. In this example, the signal does not propagate to the right due to blockage of the axon.

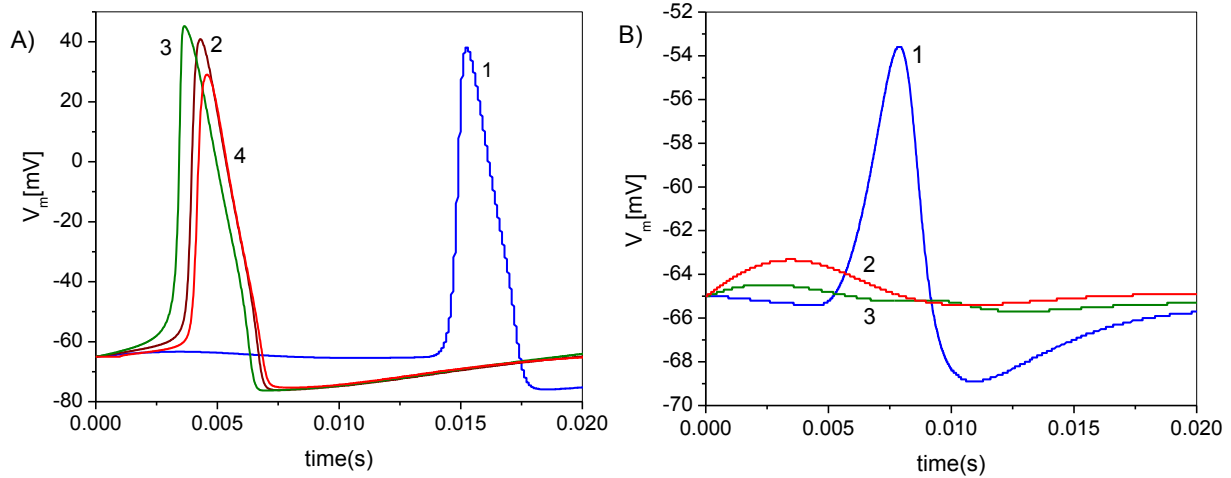

**Fig.12.** Time dependence of the action potential at various points in the axon. The excitation potential perturbation is $\delta V_m = 0$, $\xi_0 = 0.81$, $\alpha = 0$. A) Corresponds to points lying to the left of the center of the axon (where the action potential is excited), line 1 – corresponds to x=2cm, 2 – 15.5cm, 3 – 17.5 cm, 4 – 20 cm. B) Corresponds to points lying to the right of the center of the axon.1- x=22.5cm, 2 –24.5cm, 3 – 30cm.

Figure 13 shows the dependence of the coefficient $\xi_0$, at which the action potential propagation is blocked, on the size of the region where the protein channels redistribution takes place, $l_{Na}$. These results correspond to the perturbation amplitude of the resting potential $\delta V_m = 0$ (parameter $\alpha = 0$).

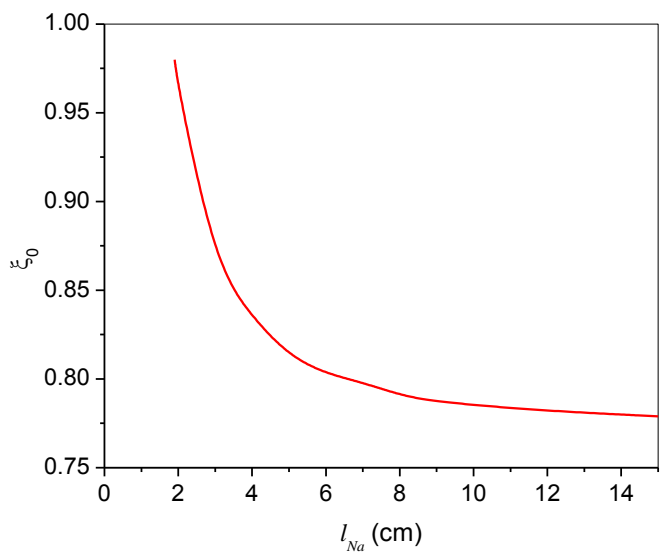

**Fig. 13.** The dependence of the coefficient $\xi_0$ in the squid axon on the size of the protein channels redistribution region $l_{Na}$ at $\alpha = 0$

It is seen that as the size of the region with redistribution of protein channels density $l_{Na}$ decreases, the value of $\xi_0$, at which the blockage of the action potential occurs, increases. And, at $l_{Na} \leq 1.8$ cm, there is no blockage, because the size of the area with ion channels depleted is larger than that on which the action potential falls, $l_{Na} > l_x$.

## IV. The possibility of an experiment on the squid axon.

As described above, the transmembrane ion channels redistribution, caused by ultrasound, can lead to the initiation of the action potential propagation or its blocking. Let us discuss some possible experiments to detect this effect in squid axon.

### A) Traveling ultrasound wave along axon.

The size of the region with depleted protein channel density is of the order $l_{Na}/2 \approx 1\text{-}4$ cm. (Fig. 14). For the traveling ultrasonic waves, this condition means that the reasonable attenuation lengths $a_d/2 \approx$ of 1-3 cm correspond to ultrasonic frequency of the order of 20-70 MHz and wave numbers $k = 2\pi f/c_s \approx 0.8 \cdot 10^5 - 2 \cdot 10^5 \text{m}^{-1}$, as follows from Fig. 6. Let us estimate power of the ultrasonic wave necessary for reorganization of transmebrane channels. It follows from (3) and (9):

$$\frac{U}{k_B T} \approx \frac{F_t a_d}{k_B T} = \frac{4\pi}{9} \frac{a^6 k^4 a_d}{k_B T} \frac{I_s}{c_s} \approx 1. \tag{29}$$

Here, we take into account that $\overline{E} = I_s/c$. That is

$$I_s \approx \frac{9}{4\pi} \frac{k_B T \cdot c_s}{a^6 a_d} \left(\frac{c_s}{2\pi f}\right)^4 \approx 1.57 \cdot 10^{13} \text{ Watt/m}^2. \tag{30}$$

In (30), the following parameters are accepted: $a = 5 \cdot 10^{-9}$ m, $a_d = 10^{-2}$ m, $c_s = 1450$ m/s, $f = 5 \cdot 10^7$ Hz.

Thus, it is not possible to observe the effect of the transmembrane channels density redistribution in the field of the ultrasonic traveling waves in the squid axon, because the irreversible change in the membrane properties, related to the membrane heating and the cavitation development in the surrounding liquid, are occurring at much lower intensities of ultrasound.

### B) Standing ultrasound wave along axon.

For the case of a standing wave, we are not limited by the frequency, because the size of the area with the channel density depletion is determined by the wave length of the ultrasound $\lambda/2 > l_{Na}$, $\lambda \gg a_d$. In accordance with (13), the condition for the action potential blocking, similar to (29), has the form:

$$\frac{\pi}{3} \cdot \frac{a^3 \overline{E}}{k_B T} \approx 1 \qquad (31)$$

It follows from (31), $\overline{E} \approx \frac{3k_B T}{\pi a^3} \approx 3.162 \cdot 10^4$ Joule/m$^3$. The intensity of the sound waves: $I_s = \overline{E} c_s \approx 4.7 \cdot 10^7$ Watt/m$^2$

Although the power at which the effect of transmembrane channels density redistribution becomes observable is 5 orders of magnitude smaller than in the traveling wave, it is still large enough for cavitation and overheating effects to be dominant.

**C) Microwave with electric field along axon.**

In [19], it is shown that the surface charges of the ions, 392 which determines the resting potential, are negative (but not 393 equal) on both surfaces of membrane and firmly binds with 394 the membrane. The electric part of the electromagnetic wave 395 acting on the charges shifts the surface of the membrane, and 396 that leads to the forced vibrations of the membrane (Fig. 14).

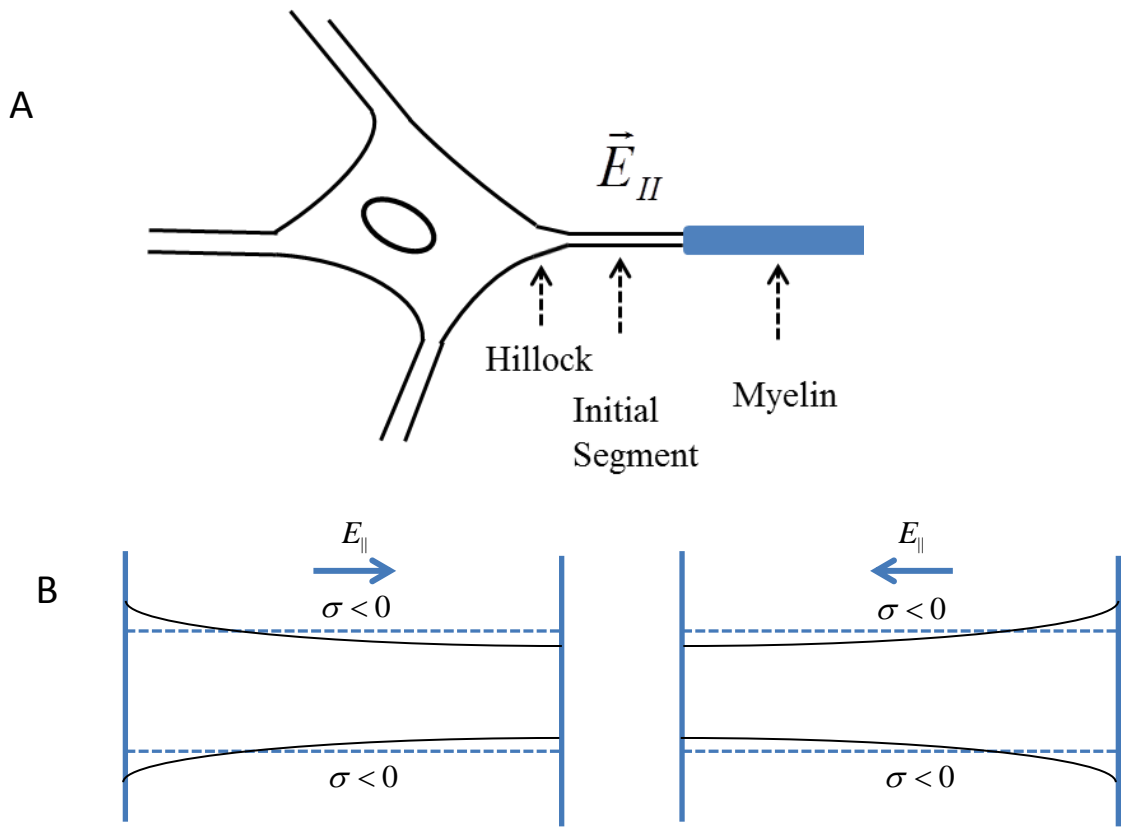

**Fig. 14.** A): Scheme of a neuron with the initial segment, hillock and the first myelin, interacting with the tangential electric field of a microwave. B): Membrane deformations cause by the microwave electric field parallel to the surface of membrane. σ is the surface charge density. Figure B) corresponds to the initial segment.

In [8], it is shown that in the range of about ~100 GHz, strongly pronounced resonances associated with the excitation of forced ultrasonic vibrations in the membrane are formed as a result of the charged membrane interacting with electromagnetic radiation (Fig. 14A). These forced vibrations create acoustic pressure, which may lead to the redistribution of the protein transmembrane channels, thus changing the threshold of the action potential excitation in the axons of the neural network. In [8], the interaction of electromagnetic microwave radiation with the initial segment – the area between the neuron hillock and the first part of the axon covered with the myelin layer was considered. Following [8], the expression for the amplitude of the longitudinal and transverse displacement of the membrane under the influence of the microwave:

$$A_{\parallel} = \frac{2\sqrt{2}}{\pi^2} \frac{\sigma}{\omega_r \nu h \rho_0 \sqrt{\varepsilon_0 \varepsilon_m^{1/2} c}} l^2 \sqrt{I_{MW}} \qquad (32)$$

$$A_{\perp} = -\frac{c_s k}{\omega_r} A_{\parallel} = -\frac{h}{l} A_{\parallel} \qquad (33)$$

Here: $I_{MW}$ is intensity of microwave, $c$ - speed of light, $\varepsilon_0$ is dielectric permittivity of vacuum, $\varepsilon_m$ - permittivity of membrane, $\sigma$ - surface charge of membrane, $\nu$ - is the effective kinematics viscosity (details see in [8]), $l$ is the size of axon part interacting with the longitudinal electric field component of the microwave; $\omega_r = \pi c_s / h$ is the resonant frequency of the microwave. Since the diameter of the axon is much smaller than the microwave wave length, $d_{axon} \approx 0.5$mm $\ll \lambda = 2\pi c / \omega_r \sim 1-10$ mm, we can assume that the electric field acting on the neuron depends only on time. Since the longitudinal displacement of the membrane $A_{\parallel}$ should be much less than $l$, and the transversal displacement, $A_{\perp}$, is much less than $h$,

$$K_I = \frac{A_{\parallel}}{l} = \frac{A_{\perp}}{h} = \frac{2\sqrt{2}}{\pi^3} \frac{\sigma}{\nu c_s \rho_0 \sqrt{\varepsilon_0 \varepsilon_m^{1/2} c}} l \sqrt{I_{MW}} \ll 1 \qquad (34)$$

The constraint (34) imposes a condition on the microwave intensity. Substituting (32) in the expression for $\overline{E}$, we obtain:

$$\overline{E} = \frac{1}{2} \rho_0 \omega_r^2 A_{\parallel}^2 = \frac{4\sigma^2}{\pi^4 h^2 \nu^2 \rho_0 \varepsilon_0 \varepsilon_m^{1/2} c} l^4 I_{MW} \qquad (35)$$

Substituting (35) into (4) we obtain the dependencies on $l$ and $I_{MW}$ of the longitudinal force acting on the ion transmembrane channel

$$F_s = \frac{8h\sigma^2}{3\pi^2 \varepsilon_0 \varepsilon_m^{1/2} c \rho_0 \nu^2} l^3 I_{MW} \sin(2\pi x/l) \qquad (36)$$

Substituting (36) into (10), we obtain:

$$\frac{U}{k_B T} \approx \frac{4h}{3\pi^3 \varepsilon_0 \varepsilon_m^{1/2} c \rho_0 \nu^2 k_B T} \sigma^2 l^4 I_{MW} \qquad (37)$$

Let us estimate the redistribution time under the influence of the longitudinal force $F_s$, taking into account the lateral mobility (6):

$$\Delta t_s \approx \frac{l}{2V_d} \approx \frac{l}{2D_L} \frac{k_B T}{F_s} \approx \frac{l^2}{4D_L} \frac{k_B T}{U} = \frac{3\pi^3 \varepsilon_0 \varepsilon_m^{1/2} c\rho_0 v^2 k_B T}{16 D_L h \sigma^2} \frac{1}{l^2 I_{MW}} \tag{38}$$

Substituting in (38) $\sigma \approx 1.24 \cdot 10^{-4}$ C/m$^2$, corresponding to the resting membrane potential of $\approx 70$ mV, $h = 5 \cdot 10^{-9}$ m, $D_L = 10^{-13}$ m$^2$/s, $\rho_0 = 9 \cdot 10^0$ kg/m$^3$, $\varepsilon_m = 2$, $v \approx 6.3 \cdot 10^{-8}$ m$^2$/s, $T = 300^\circ$ K, at assumed longitudinal sound velocity, $c_s = 1500$ m/s, we obtain:

$$f_r = \frac{\omega_r}{2\pi} = \frac{c_s}{2h} = 150 \text{ GHz} \tag{39}$$

$$K_I = 1.8 \cdot 10^3 l \sqrt{I_{MW}} \tag{40}$$

$$\Delta t_s \approx 1.3 \cdot 10^{-5} \frac{1}{l^2 \cdot I_{MW}} \tag{41}$$

Consider, as already mentioned, the experiments with a squid axon as model experiments for the processes in the initial segment of the neuron with myelined axon. Taking, for example, $l = 0.05$ m as a length for the interaction region with the resonant microwave radiation with intensity $I_M = 1$ Watt/m$^2$, we obtain: $K_I = 0.9 \cdot 10^{-4}$, $\Delta t_s \approx 5.4 \cdot 10^{-3}$ s. For a real initial segment length of $l \approx 50 \mu m$ at the same parameters of the membrane and the characteristics of microwave radiation, we obtain, respectively, $K_I = 0.09$ (ie, elongation of the membrane is small enough to have been valid for the linear elasticity theory), $\Delta t_s \approx 5.4 \cdot 10^3$ s.

**V. Discussion**

As shown in this study, the redistribution of transmembrane channels, caused by standing or traveling ultrasonic waves in the membrane of the initial segment, leads initially to a decrease in the excitation threshold of nerve impulses, and then, with increasing amplitude of the channels density perturbations, to the possibility of spontaneous excitation of the action potential. However, the effect of purely mechanical ultrasonic vibrations is not as effective as the action of the forced electromechanical oscillation at the resonant frequencies, excited by the interaction of charged membranes with the microwave. Further increases in the amplitude of the transmembrane channel density perturbations can block the propagation of a pulse in areas with rarefied density of channels. This effect was likely observed in experiments [20], where the frog nerves were irradiated for 70 seconds with the microwave pulse of the frequency 60.125 GHz at a very weak intensity ~70 - 770 nWatt/cm$^2$. At the minimum microwave power, a spontaneous excitation of the action potential was observed. With increasing the microwave power, the spontaneous excitation stopped and occurred again after switching off power. These results are fully qualitatively consistent with our model. In [20], a change in the shape of the action potential and shortening of the duration of the refractive period with increasing of the microwave power was also observed. All these experimental effects may be associated with the density variation of the transmembrane ion channels, induced by ultrasound [8]. It should be noted that the microwave frequency 60.125 GHz, used in [20], is likely to be close to the resonance frequencies of longitudinal vibrations of the axon membrane [8].

It has to be mentioned that for the effective interaction of charged particles attached to the membrane with the external field, the value of the field may be small. Regardless of the magnitude of the field, the energy and the momentum are transferred from the field to the charged particles. Thus, the random thermal motion is superimposed on a small ordered motion. This occurs, for example, in an electrolyte (plasma), when the drift velocity of the charged particles is a million times less than their average thermal velocities. The microwave field is interacting with the charges on the membrane and transmits the momentum to them, which, in turn, is transmitted to the membrane and causes forced oscillations. The same holds for a weak ultrasonic wave interacting with suspended particles (protein channels). As a result, an acoustic drift appears, superimposed on the average thermal motion, even at an ultrasound with a very low intensity. In this case, the potential of the acoustic force acting on the particles can be much smaller than the thermal energy, as it happens in the hydrodynamic or gas flows where the directional velocity is much smaller than the averaged thermal velocity, and the corresponding pressure drop is much lower than the unperturbed static pressure.

It should be noted that the resonant frequency (39), at which the displacement of the membrane is maximal, is proportional to the speed of sound in the membrane. In our estimations, we chose $c_s = 1500$ m/s, equal to the velocity of sound in the oil. It is difficult to know the exact value of the longitudinal speed of sound in the membrane at high frequencies, which could be much lower [21]. Since the resonance is very narrow [4], the experimentally measured frequencies, at which there is a blocking or self-excitation of the action potential, enable the determination of the velocity of sound in the membrane.

It appears that the best experimental methods to observe the effect of transmembrane channel redistribution described in this article and in [8], correspond to the measurements of the spatial distribution of sodium channels along the axon membrane, including electro-optical measurements (see, [22, 23]), since they allow one to directly measure the density distribution of the transmembrane channels prior to the ultrasound (or microwave) pulse and immediately after. An indirect confirmation of transmembrane channel redistribution would be the observation of the action potential excitation and propagation (or blocking) dependence on the amplitude of the ultrasound. This can be done using conventional probe techniques [24], as well as non-intrusive optical methods; e.g. observing the dynamics of the second-harmonic generation of laser radiation scattered by the nerve fibers[25,26].

**Conclusions:**

1. The standing and traveling ultrasonic waves can lead to a redistribution of transmembrane protein channels in the membrane of the axon initial segment, with the formation of regions of high or low surface densities of channels with respect to their uniform equilibrium surface distribution. However, the required intensity of ultrasound is very high, so that thermal and cavitation effects become dominant. On the other hand, the forced resonant electromechanical oscillations induced by the action of the longitudinal component of the microwave electric field may lead to the essential redistribution of transmembrane channels without significant thermal effects.
2. The redistribution of transmembrane ion channels leads to the decrease in the beginning of the excitation threshold of the nerve impulse, and then, with an increase in the amplitude of density perturbations, to the possibility of spontaneous excitation of the action potential. Further increase in the amplitude of the transmembrane channels density perturbation may cause blocking of the pulse in the region with depleted channel density.

3. The blocking effect is dependent on the width of the area with a low density of channels. With the narrowing of the area of depleted transmembrane channels density, the blocking is less effective, and, at width of the order of the action potential scale or less, the blocking is not possible.
4. The effect of blocking of the action potential can be used for anesthesia. Its advantage over conventional ultrasonic method of anesthesia is that it does not violate the integrity of the membranes of axons and is reversible. After the source of ultrasound is turned off, the equilibrium density of transmembrane channels is restored as a result of the lateral diffusion.
5. In the paper, we considered the effect of ultrasound on transmembrane sodium channel; however, the same effect of ultrasound remains true for other transmembrane ion channels, which are subject to lateral diffusion.